# Suppressed out-of-plane polarizability of free excitons in monolayer $WSe_2$

Ivan A. Verzhbitskiy[1,2,*], Daniele Vella[1,2], Kenji Watanabe[3], Takashi Taniguchi[3] and Goki Eda[1,2,4,*]

[1] Department of Physics, National University of Singapore, Singapore
[2] Centre for Advanced 2D Materials, National University of Singapore, Singapore
[3] National Institute for Material Science, 1-1 Namiki, Tsukuba 305-0044, Japan
[4] Department of Chemistry, National University of Singapore, Singapore
*Corresponding author: ivan@nus.edu.sg (IV); g.eda@nus.edu.sg (GE)

## ABSTRACT

Monolayer semiconductors are atomically thin quantum wells with strong confinement of electrons in two-dimensional (2D) plane. Here, we experimentally study the out-of-plane polarizability of excitons in hBN-encapsulated monolayer $WSe_2$ in strong electric fields of up to 1.6 V/nm (16 MV/cm). We monitor free and bound exciton photoluminescence peaks with increasing electric fields at a constant carrier density, carefully compensating for unintentional photodoping in our double-gated device at 4K. We show that the Stark shift is $< 0.4$ meV despite the large electric fields applied, yielding an upper limit of polarizability $\alpha_z$ to be $\sim 10^{-11}$ Dm/V. Such a small polarizability, which is nearly two orders of magnitude smaller than the previously reported value for $MoS_2$, indicates strong atomic confinement of electrons in this 2D system and highlights the unusual robustness of free excitons against surface potential fluctuations.

Quantum-confined Stark effect (QCSE), also known as giant Stark effect, refers to the electric-field-induced change in the optical transition energies of quantum electronic systems due to spatial separation of electron and hole wavefunctions [1]. It is commonly observed in quantum dots and wells, and serves as the working principle of some electro-optical modulators. The ability to electrically tune exciton resonance energies by QCSE is attractive for novel photonic devices such as polaritons transistors [2] and tunable quantum light sources [3]. On the other hand, QCSE due to interfacial built-in fields and random potential fluctuations leads to undesirable shift in the emission wavelength of light-emitting devices, imposing a fundamental limit on device performance [4].

Two-dimensional (2D) semiconductors such as monolayer transition metal dichalcogenides (TMDs) are a unique system to study QCSE due to the molecular-scale confinement of excitons along the c-axis. In such systems, the out-of-plane exciton polarizability $\alpha_z$, which determines the magnitude of the Stark Shift according to $\Delta E \sim -\alpha_z F_z$ where $F_z$ is the electric field, is expected to be small. Conversely, measurement of polarizability can provide useful insight into the atomic confinement potential of electrons and holes in these systems.

Accurate measurement of Stark shift in monolayer TMDs is challenging due to the intrinsically small polarizability, which is often masked by the strong shift arising from charge modulation. High quality interface is therefore crucial in minimizing trapping of photocarriers, which leads to unintentional photodoping (or photo-gating) and corresponding apparent field-induced shift [5]. Roch *et al*. [6] addressed this issue by identifying the bias window within which photodoping did not occur and limiting the electric field to below 0.17 V/nm (1.7 MV/cm) to observe Stark shift. They observed a small quadratic shift of up to ~0.4 meV for both exciton and trion in monolayer $MoS_2$ encapsulated in hexagonal boron nitride (hBN), and determined their out-of-plane polarizability to be 7.8 x $10^{-10}$ D m $V^{-1}$, which is an order of magnitude smaller than the estimation in the initial work [5]. While this value is apparently consistent with the theoretical prediction based on infinite square well model, it is significantly smaller than that based on finite square model [7]. Chakraborty et al. [3, 8] conducted similar low-field studies on localized excitons in hBN-encapsulated monolayer $WSe_2$, and observed larger Stark shift of 2 ~ 20 meV. Bilayer $WSe_2$ has shown to exhibit significantly enhanced Stark shift due to the layer-polarized excitons rather than due to intralayer polarizability of excitons [9]. The polarizability of free exciton species in monolayer TMDs remains a curious problem especially in the high field regime where polarizability is saturated and quadratic shifts are no longer expected.

In this Article, we report on the Stark shift of free and charged excitons in double-gated monolayer $WSe_2$ up to high electric fields of 1.6 V/nm (16 MV/cm), which is nearly one order of magnitude greater than that in the previous study by Roch *et al*. [6]. We demonstrate that high field measurements can be conducted without the influence of photodoping by carefully compensating the excess doping with suitable gate biases. We show, contrary to the earlier reports, Stark shift is < 0.4 meV despite the significantly larger electric fields, which yields an upper limit of $\alpha_z$ to be ~ $10^{-11}$ Dm $V^{-1}$, nearly two orders of magnitude smaller than that reported for $MoS_2$ [6].

## EXPERIMENTS

Bulk $WSe_2$ crystals were grown by direct vapor transport (DVT) or flux method. W powders (purity: 99.999%) and Se shot (purity: 99.999%) were sealed inside a quartz ampoule at a ratio of 1:20 at ~$10^{-4}$ Torr. The ampoule was then transferred into the annealing furnace where it was kept at 1000 °C for 2 days and then slowly cooled to 400 °C with rate of 2 °C/hour. Upon reaching 400 °C, the furnace was switched off allowing the ampoule to reach room temperature naturally. Small single-crystalline flakes of $WSe_2$ were then harvested from selenium powder built-up in the ampoule. Monolayer samples were exfoliated by mechanical exfoliation and sandwiched by hBN and few-layer graphene (FLG) flakes using dry-transfer and pick-up technique [10], forming FLG/hBN/$WSe_2$/hBN/FLG heterostructures. Electrical contacts for $WSe_2$ and FLG layers were fabricated using laser lithography and thermal evaporation.

## RESULTS

Figure 1a and b show the schematic and optical image of our double-gate device consisting of hBN/$WSe_2$/hBN quantum well sandwiched between FLG layers that serve as the electrodes. The $WSe_2$ layer is in contact with FLG flake and grounded Au electrode to allow independent control of the top ($V_{TG}$) and bottom ($V_{BG}$) biases. Figure 1c shows the low temperature photoluminescence spectrum of a typical sample without electrical bias. The ground free exciton emission line is most frequently observed around 1.716 eV, typically having a width of 5~7 meV. It is worth nothing that we did not observe dark excitons with our NA0.8 objective [11]. Some samples exhibited numerous emission peaks below the ground free exciton peak due to defect-related localized states [12]. Here we focus on high quality samples that showed predominantly free exciton emission features with a narrow linewidth (~ 5 meV).

We first start by examining the doping-induced changes in the emission spectrum. Figure 2a shows 2D color map showing the bottom-gate ($V_{BG}$) dependence of photoluminescence intensity where the top FLG layer is at a floating potential. This typical gated photoluminescence spectrum shows the emergence of negatively ($X^-$) and positively ($X^+$) charged exciton peaks with increasing positive or negative gate biases. Representative emission spectra in different doping regimes are shown in Figure 2b. At a moderate doping level we observe splitting of negative trion into intra- and inter-valley species due to short-range electron-electron exchange interactions [13]. The associated binding energies of positive ($X^+$) and two negative trions ($X^-_1$ and $X^-_2$) are 22 meV, 31 meV, and 37 meV, respectively, in agreement with the previous reports [13, 14]. The general trends in doping-induced changes in peak energy (Fig. 2c) linewidth and intensity (Fig. 2d) are also consistent with the reported behaviors [13, 15-17]. We use these results as a reference to identify any unintentional photodoping during double-gated photoluminescence measurements.

In a double-gated device, the carrier density and electric field can be independently controlled. Carrier density $n$ is determined by $|V_{TG} + V_{BG}|$ whereas the out-of-plane field strength $E$ scales with $|V_{TG} - V_{BG}|$. Here we assume the capacitance of the top and bottom hBN layers to be equal. In the plot of $V_{TG}$ vs $V_{BG}$, the carrier density is therefore constant along the line defined by $|V_{TG} - V_{BG}|$ = const, and conversely, the electric field strength is constant along the $|V_{BG} + V_{TG}|$ = const line (see Fig. 3a). Thus, we expect to observe constant ($I(X^{+/-})/I(X^0)$) along the equi-density line because the intensity ratio is solely determined by mass-action law [18]. However, we find that the intensity ratio continues to drift as we conduct the measurements (see Fig. 3b), which is a signature of photo-doping arising from trapping of photo-excited carriers. Normally, additional negative gate bias needs to be applied to compensate for photodoping, which implies that the photoexcited holes are trapped, leaving behind free electrons. To investigate QCSE, we experimentally determined the equi-density condition by controlling $V_{BG}$ and $V_{TG}$ such that the $I(X^{+/-})/I(X^0)$ remained constant within an error of ±1.5% and the electric field was calculated for each condition where photoluminescence with the desired $I(X^{+/-})/I(X^0)$ was measured. Figure 3c shows the experimentally derived equi-density line (symbols) against the geometrically predicted trend (dashed line). We extracted the density of photo-induced trap carriers as a function of displacement field and plotted it in Figure 3d. It reveals that the density of photo-activated traps increases linearly with electric field, reaching > $10^{12}$ $cm^{-2}$ in the high field regime.

We thus obtained the photoluminescence spectrum at different out-of-plane electric field keeping the carrier density constant. Figure 4 shows changes in the energy and linewidth of free exciton ($X^0$) and two species of negative trions ($X_1^-$, $X_2^-$) as a function of electric field for low ($n < 10^{10}$ $cm^{-2}$), and two intermediate electron doping levels ($n \sim 8 \times 10^{11}$ $cm^{-2}$ and $n \sim 10^{12}$ $cm^{-2}$). Interestingly, the free exciton and trion energies virtually remain constant with field for all carrier densities. Taking the maximum shift to be the uncertainty of 0.4 meV, we determine the upper limit of $\alpha_z$ to be 8.8 x $10^{-12}$ Dm/V for both neutral excitons and trions. This value is nearly two orders of magnitude smaller than that reported for similar hBN-encapsulated $MoS_2$ [6] and the theoretical prediction based on infinite square well model ($\alpha_\perp^\infty \sim 9.6 \times 10^{-10}$ Dm/V) [7].

While QCSE is effectively absent for free excitons and trions, the neutral exciton peaks exhibit consistent broadening $\Delta\Gamma$ with increasing field. This broadening may be attributed to field-activated dephasing, recombination, or dissociation. We estimate the time scale of this new channel to be on the order of 1.2 ps from the relation $\tau = \hbar/\Delta\Gamma$. Since the electric field is sufficiently large to overcome the Coulomb binding energy of the excitons, the broadening may be explained by field-dissociation through tunneling. However, we do not observe consistent reduction in the intensity of emission. Considering that trions, which are more susceptible to field-dissociation than neutral excitons, do not show the same trends, the broadening may be attributed to dephasing due to field-enhanced inhomogeneity rather than dissociation.

Now we consider the origin of remarkably small out-of-plane polarizability of excitons and trions in our $WSe_2$ device and difference with the previously studied $MoS_2$. Infinite square model predicts that the polarizability scales with the well thickness $d$ as $\alpha_\perp^\infty \sim d^4$ [7]. While the initial prediction assumed a well width of $d \sim 6.5$ Å, our results indicate that the effective well width is at least ~32 % of the total layer thickness. That is, the electrons are tightly confined in the atomic orbitals of the ions. Given the similar atomic structure and thickness of monolayer $WSe_2$ and $MoS_2$, it is reasonable to expect similar spatial degree of freedom and therefore polarizability for electrons and holes in both materials. However, when encapsulated in hBN, band edge carriers are subject to different confinement potentials in the two heterostructures. Figure 5 shows the band alignment of hBN/$WSe_2$/hBN and hBN/$MoS_2$/hBN heterostructures based on DFT calculations [19, 20]. For both materials, electrons experience strong confinement barrier ($\Delta_e > 2$ eV) from the conduction band offset $\Delta_e = E_c^{TMD} - E_c^{hBN}$. The holes, on the other hand, are significantly less confined due to smaller valence band offset. While exact band offset for TMDs and hBN is still under debates [20-22], difference in valence band maxima of $WSe_2$ and $MoS_2$ (~ 1 eV) implies that the holes in $WSe_2$ are confined deeper inside the hBN's bandgap ($\Delta_h^{WSe2} > 1.5$ eV, $\Delta_h^{MoS2} < 1.0$ eV). Given that the exciton polarizability is strongly dependent on the quantum well depth [7], the robust in-plane stability of excitons in hBN/$WSe_2$/hBN can be understood as resulting from the large confinement energy barrier.

In summary, we have conducted Stark shift measurement of free exciton and trion in hBN-encapsulated monolayer $WSe_2$ up to a high electric field of 1.6 V/nm. Our double-gated spectroscopy shows that van der Waals heterostructures are highly susceptible to significant photodoping density of ~ $10^{12}$ $cm^{-2}$ in high field conditions. We have demonstrated that the electron motion is strongly suppressed in the out-of-plane direction with exciton polarizability nearly two orders of magnitude smaller than that of $MoS_2$. We attribute the striking difference in the polarizability of these two similar materials to the different confinement potential due to hBN. Our observation suggests that smaller gap TMDs with larger electron affinity such as $MoTe_2$ also exhibit negligible Stark shift when sandwiched with hBN due to large confinement energy barriers for both types of carriers.

## METHODS

### Double-gated photoluminescence spectroscopy

The device was measured in continuous He-flow cryostat (Oxford Hi-Res2) at base temperature of 4 K. Top and bottom gate biases were sourced by Agilent B2902A high precision source/measure unit, while $WSe_2$ flake was set to ground. Top and bottom gate leakage currents were continuously monitored to stay below 1nA limit. Electric field was estimated as [23, 24] $F_\perp = (F_{\mathrm{TG}} - F_{\mathrm{BG}})/2$, where $F_{\mathrm{TG}} = \varepsilon_{\mathrm{TG}}(V_{\mathrm{TG}} - V_{\mathrm{TG}}^0)/d_{\mathrm{TG}}$ and $F_{\mathrm{BG}} = \varepsilon_{\mathrm{BG}}(V_{\mathrm{BG}} - V_{\mathrm{BG}}^0)/d_{\mathrm{BG}}$ are fields independently tuned by top and bottom gate biases, respectively. Here $\varepsilon$ and $d$ are the dielectric constant and thickness of dielectric barrier and $V^0$ is the voltage bias required to compensate for initial environmental doping.

Photoluminescence was excited with 532 nm (2.33 eV) laser focused on the sample with long-working distance 100x objective (NA0.8) with power < 5μW. Emitted light was collected with the same objective lens (confocal geometry) and analyzed by NT-MDT custom-built spectrometer equipped with CCD camera. PL spectra were fitted with multi-Lorentz model and peak position, width (FWHM), and area values were extracted. Experimental uncertainty takes into account spectral resolution, fitting errors and standard deviation for repeatedly measured spectra.

**ACKNOWLEDGEMENTS**

Authors would like to thank Prof. José Viana-Gomes and Prof. Thomas Garm Pedersen for the fruitful discussions. Authors acknowledge the Singapore National Research Foundation for funding the research under medium-sized centre programme. G.E. also acknowledges support from the Ministry of Education (MOE), Singapore, under AcRF Tier 2 (MOE2015-T2-2-123, MOE2017-T2-1-134) and AcRF Tier 1 (R-144-000-387-114).

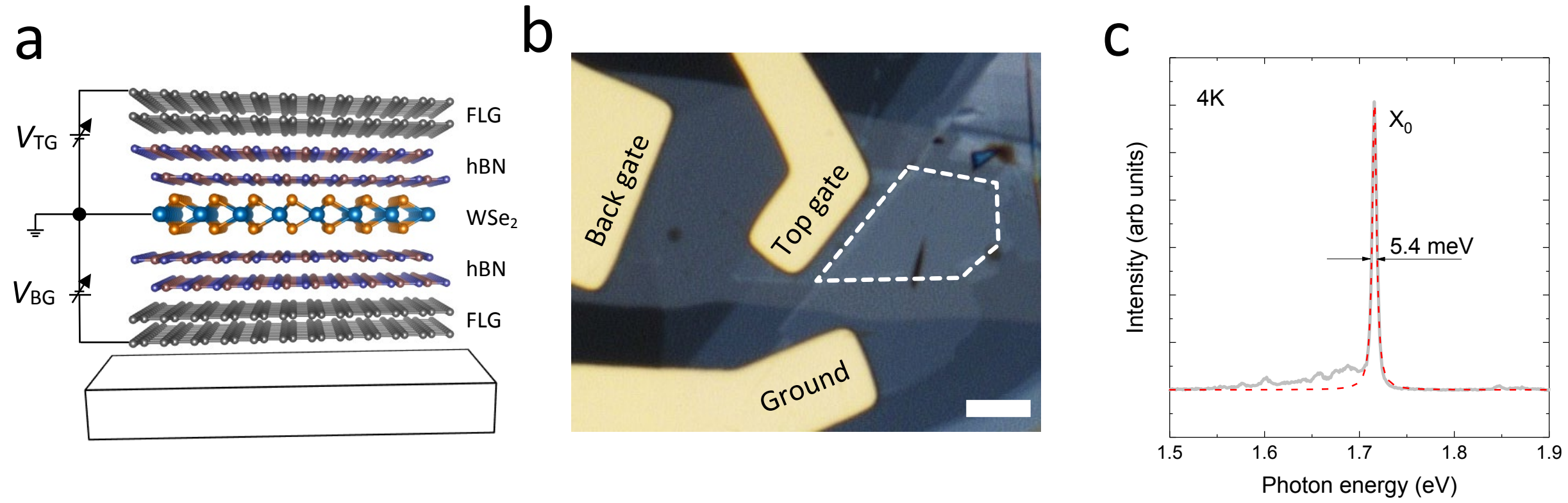

**Figure 1.** (a) Schematics of double gate device. The vertical heterostructure is formed by encapsulating monolayer $WSe_2$ in hBN and semi-transparent few-layer graphene. (b) Optical image of the heterostructure device. Dashed line highlights the overlapping area consisting of FLG/hBN/$WSe_2$/hBN/FLG stack. Scale bar is 5 μm. (c) Photoluminescence spectrum of the heterostructure at 4 K. Solid line represents experimental curve and the dashed line is the Lorentz fit for neutral $X^0$ exciton. Fit yields FWHM of 5.4 meV. Absence of prominent defect peaks at lower energies indicate the high quality of the sample.

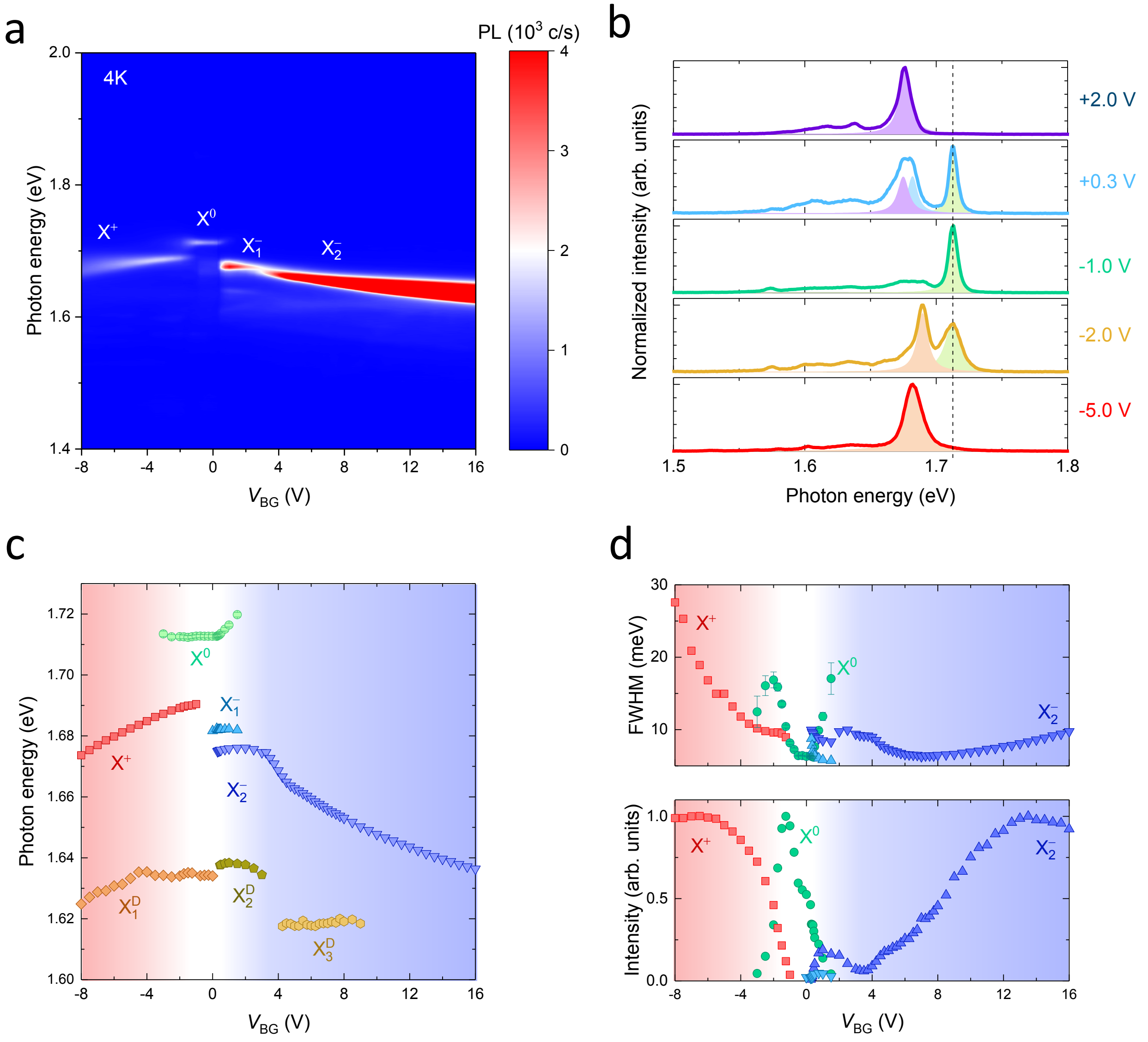

**Figure 2.** Single-gate photoluminescence spectra at 4 K. The gate bias $V_{BG}$ is applied to the bottom FLG layer while the top FLG is kept at floating potential. (a) 2D color map of photoluminescence intensity as a function $V_{BG}$. (b) Representative photoluminescence spectra for different doping regimes: moderate *n*-doping ($V_{BG}$ = 2 V); low *n*-doping ($V_{BG}$ = 0.3 V); charge neutrality ($V_{BG}$ = - 1.0 V); low *p*-doping ($V_{BG}$ = -2.0V); and moderate *p*-doping ($V_{BG}$ = -5 V). (c, d) Variation of (c) exciton peak positions, (d) linewidth (FWHM) and intensity as function of $V_{BG}$.

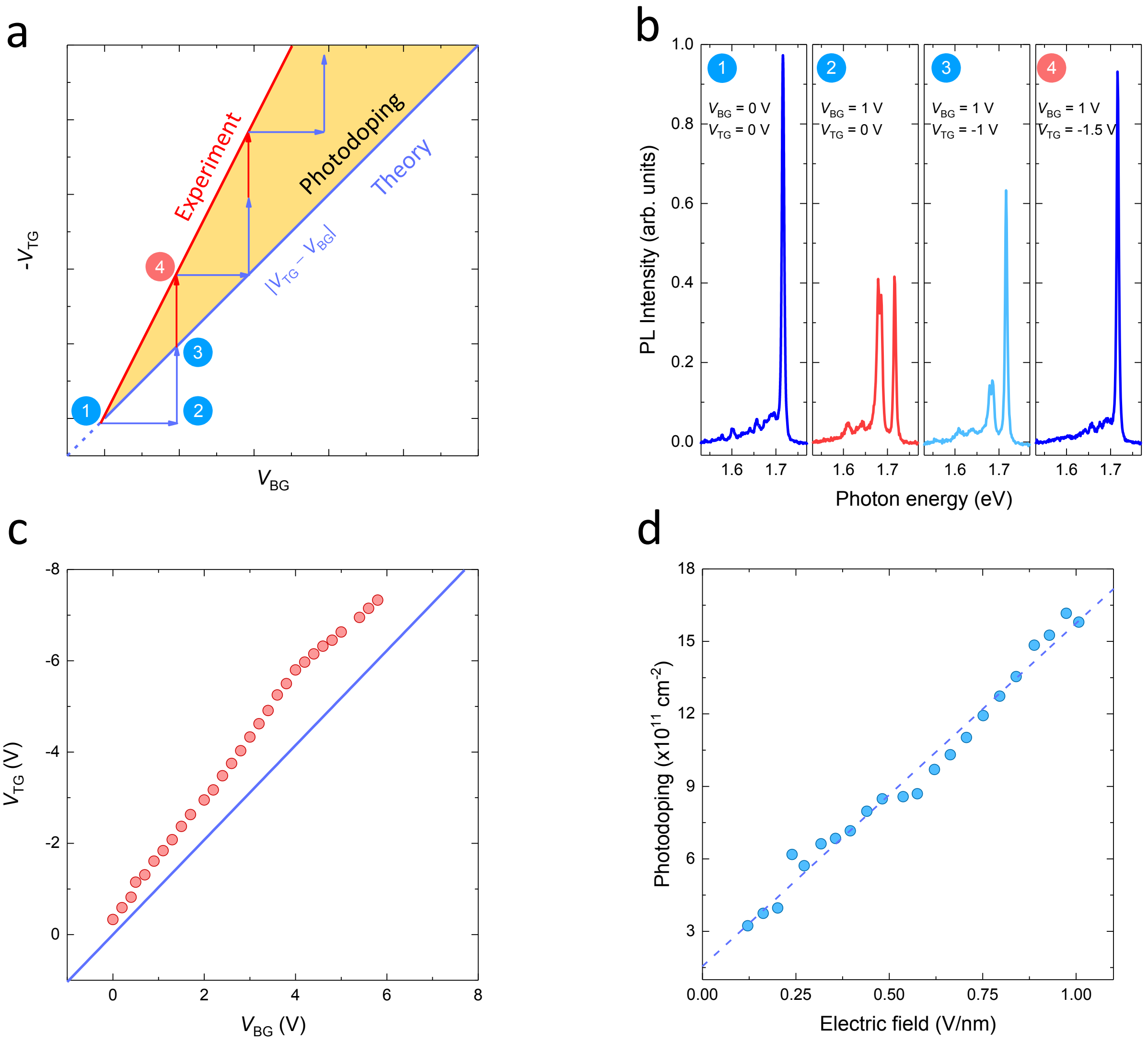

**Figure 3.** Photodoping compensation by double-gating. (a) Schematics showing the effect of photodoping on the equi-density line and its experimental determination. Blue solid line represents theoretically defined equi-density line based on the top and bottom gate capacitance. Red solid line represents the experimentally determined equi-density line, which deviates from the theoretical line due to photodoping. To determine the experimental equi-density line, photoluminescence spectrum is collected at points 1-4 (the arrows represent the path followed). To recover the initial doping density, $V_{TG}$ larger than theoretically expected is applied to compensate for photodoping (step 3-4). (b) Photoluminescence spectra recorded at each step shown in (a). Charge-neutrality condition is maintained by monitoring the intensity ratio $I(X^-)/I(X^0)$. Photoluminescence spectra shown in step 1 and step 4 are taken as equi-density (charge-neutral) emission at two different values of electric fields. (c) Comparison of theoretical (blue solid line) and experimental (symbols) equi-density conditions. (d) Extracted photodoping carrier density as a function of electric field.

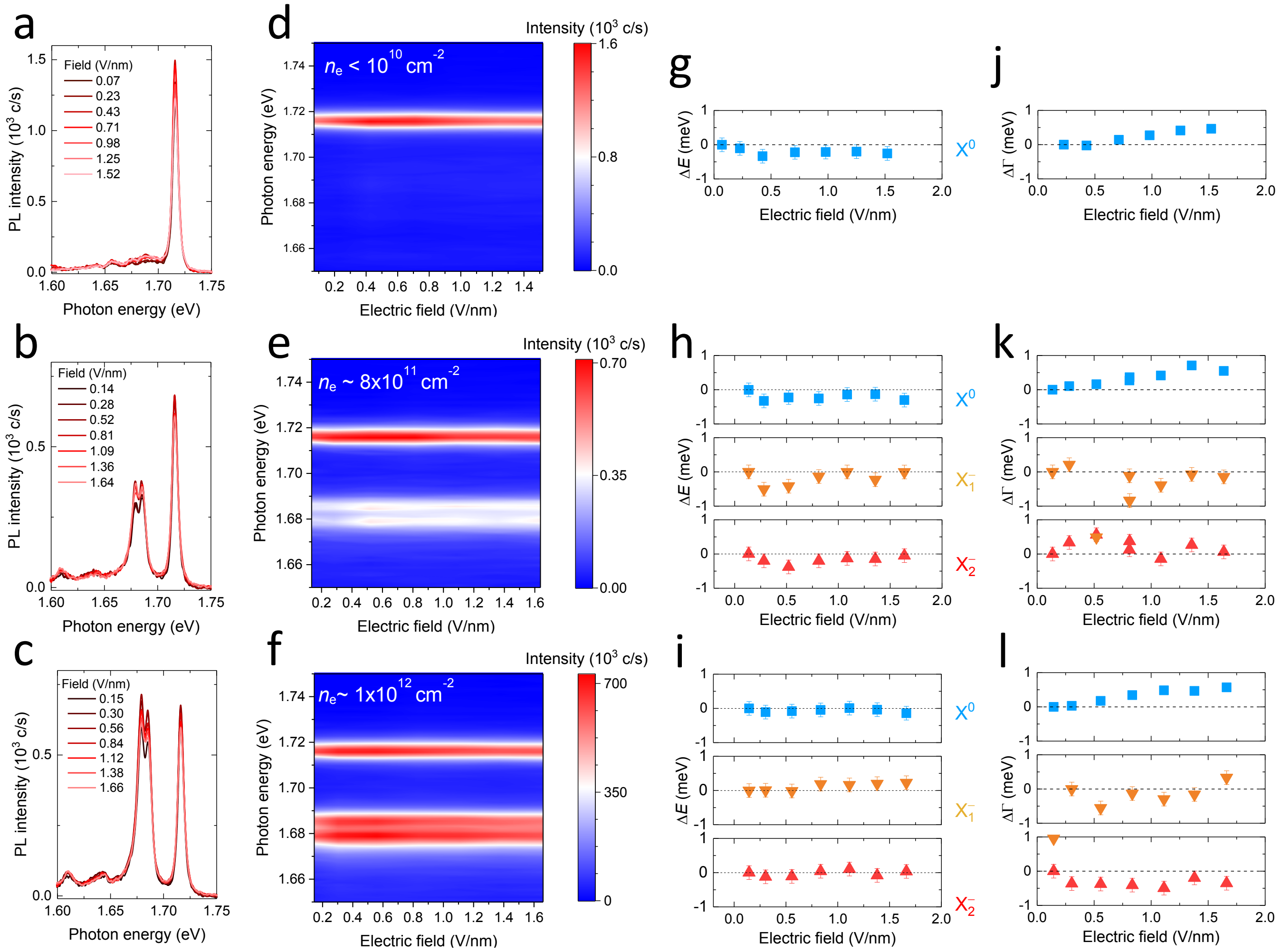

**Figure 4**. Evolution of excitonic emission features as a function of electric fields along the experimentally determined equi-density lines. (a-c) Exciton emission spectra, (d-f) emission intensity color map, (g-i) Stark shift ($\Delta E$), and (j-l) linewidth broadening ($\Delta\Gamma$) for (a, d, g, j) charge neutrality ($<10^{10}$ cm$^{-2}$), (b,e,h,k) low *n*-doping ($\sim 8\text{x}10^{11}$ cm$^{-2}$) and (c,f,i,l) moderate *n*-doping ($\sim 1\text{x}10^{12}$ cm$^{-2}$) conditions.

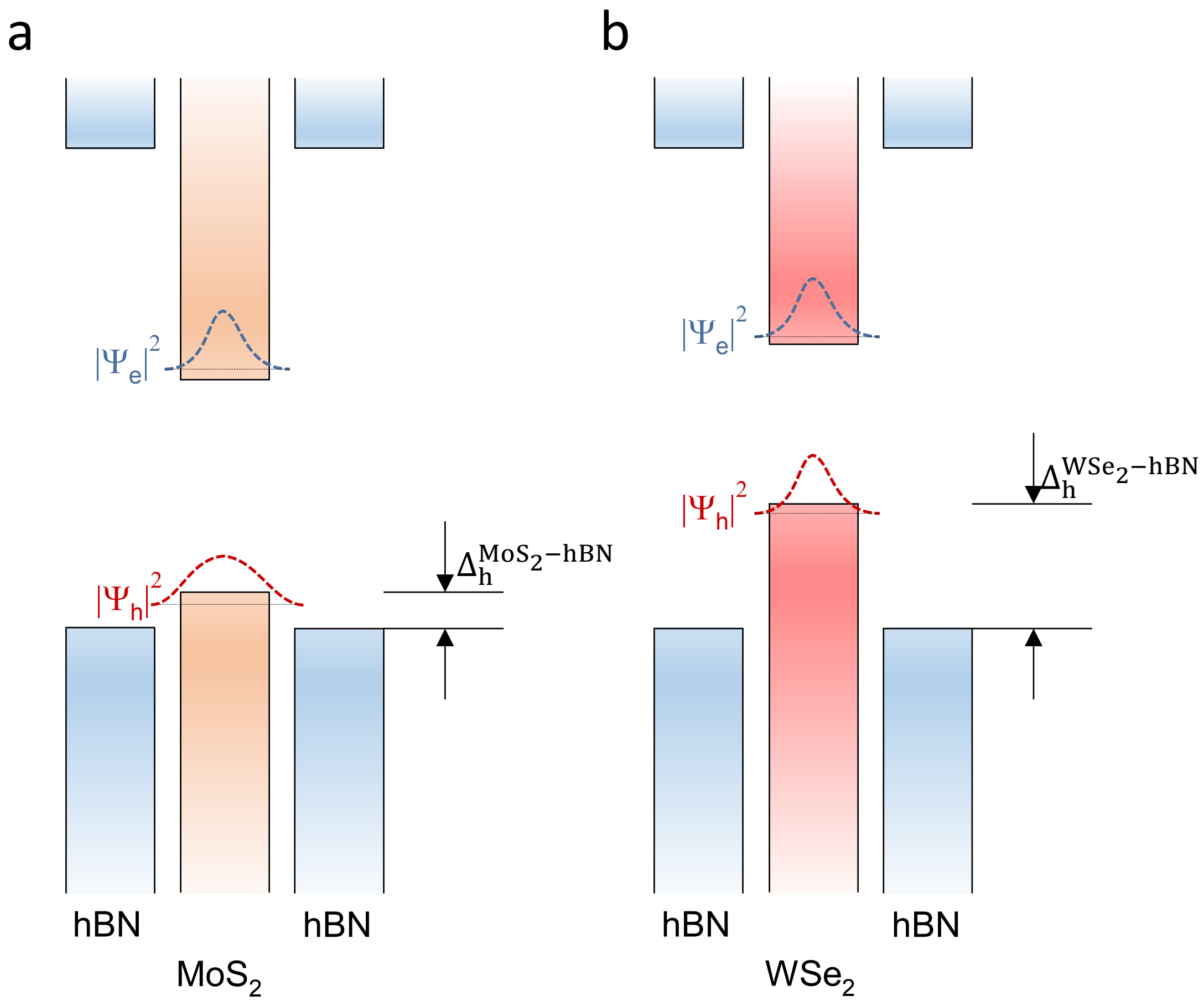

**Figure 5.** Band alignment for monolayer (a) $MoS_2$ and (b) $WSe_2$ encapsulated in hBN. Due to smaller offset of valence band maximum energies ($\Delta_h$), holes in hBN/$MoS_2$/hBN experience weaker confinement as compared to those in hBN/$WSe_2$/hBN, resulting in larger wavefunction spread and correspondingly larger polarizability.